# Depression Detection using Resting State Three-channel EEG Signal

Qiuxia Shi[1], Ang Liu[1], Rongyan Chen[1], Jian Shen[1], Qinglin Zhao*[1], Bin Hu*[1,2,3,4]

[1] Gansu Provincial Key Laboratory of Wearable Computing, School of Information Science and Engineering, Lanzhou University, China

[2] CAS Center for Excellence in Brain Science and Intelligence Technology, Shanghai Institutes for Biological Sciences, Chinese Academy of Sciences, China

[3] Joint Research Center for Cognitive Neurosensor Technology of Lanzhou University & Institute of Semiconductors, Chinese Academy of Sciences, China

[4] Open Source Software and Real-Time System (Lanzhou University), Ministry of Education, China

*corresponding author(s):Bin Hu (bh@lzu.edu.cn), Qinglin Zhao(qlzhao@lzu.edu.cn)

**Abstract.** In universal environment, a patient-friendly inexpensive method is needed to realize the early diagnosis of depression, which is believed to be an effective way to reduce the mortality of depression. The purpose of this study is only to collect EEG signal from three electrodes Fp1, Fpz and Fp2, then the linear and nonlinear features of EEG used to classify depression patients and healthy controls. The EEG recordings were carried out on a group of 18 medication-free depressive patients and 25 gender and age matched controls. In this paper, the selected features include three linear (maximum, mean and center values of the power) and three nonlinear features (correlation dimension, Renyi entropy and C0 complexity). The accuracy and effectiveness of classification model between depressive and control subjects were calculated using leave-one-out cross-validation. The experimental results indicate that selected three channel EEG and features can distinguish the subjects between depression and normal beings, the classification accuracy is 72.25%. It is hoped that the performed results can provide more choices for the early diagnosis of depression in a universal environment.

## Introduction

Depression, as a common mental illness worldwide, is predicted to become the second leading cause of disability and death before 2030[1, 2]. Depression severely affects the beings' ability to work, quality of life and interpersonal communication. It is classified as mild, moderate or severe depending on the severity of the symptoms. Nowadays, the diagnosis of depression is primarily through the clinical interviews and structured or semi-structured symptom severity scales to assess the intensity of subjective symptoms, all of which require accurate self-report from patient. Depression can be treated with psychotherapy or medication if diagnosed correctly, but it remains a chronic health problem in society as it is not diagnosed in a timely way [3].

In recent years, with the rapid development of cognitive science research technology and sensor technology, Electroencephalogram (EEG) features have been successfully applied for investigation of brain behavior in various mental diseases [4-6]. The electroencephalogram (EEG) records the electrical activity within neuronal dendritic potentials of the brain which can imply the neuronal activity [7]. EEG is easy to collect across the surface of brain, and safe, low cost, noninvasive, with high temporal resolution. It is widely used in brain function studies. Recently, many studies demonstrated the relationship between depression and EEG [8-10]. EEG power was found to be increased in central, occipital, parietal and posterior temporal areas in patients in the early stage of depression [11]. Vera et al. [12] reported that the development of depression in the parietal and occipital region of the power spectrum increased apparently. Multichannel EEG characteristics have been under consideration for investigation of depression in many studies [5, 13, 14], But more channels also mean more complexity.

The underlying neurophysiological mechanism od EEG indicates that EEG signals stem from a highly nonlinear system [15, 16]. Nonlinear complexity analysis of EEG is promising for obtaining additional information to that achieved by linear measures. Nonlinear features such as the correlation dimension (CD), Renyi entropy, C0 complexity and Lempel-Ziv complexity (LZC). G. Muralidhar Bairy et al. applied several nonlinear features, including CD, to the classification of normal EEG and depressed EEG [17]. LZC have been estimated by researchers to detect differences in psychological states and to investigate the dynamic brain mechanisms underlying the EEG. Nonlinear features of EEG signal have been shown related with emotion, intensity of brain activity, attentional deficits and so on. The Lempel-Ziv complexity (LZC) of multichannel resting EEG has been reported being successful in evaluation of various neuronal and mental disorders including major depression [18-20].

Due to the brain is a complex system of linear and nonlinear combinations, the present study aimed at extending previous findings and using a combination of linear and nonlinear features extracted from the EEG to differentiate normal and depression EEG signals. Our studies select three number of electrodes in the 10-20 system, Fp1, Fpz and Fp2 respectively. Based on previous research results of linear features and nonlinear features, we use SVM classifier to implement dichotomy. Through analysis, we can distinguish the depression from normal populations, the highest accuracy was 72.25%, Of course, due to the complexity of the generation of EEG signals, it is difficult at present to produce applicable universally research results.

## Materials and Methods

### *Participants*:

In our study, 26 persons (11 female and 15 male) diagnosed with depression and 28 (10 female and 18 male) demographically similar normal controls were recruited from the psychiatric hospital. All subjects are medication free and right handedness. Patients group and control group are matched in age (between 18 and 53), gender and handedness. The subjects were asked to restrain themselves oneself from coffee for two hours and from alcohol for 24 hours before the experiment.

In addition, the included patients did not have major medical or neurological illnesses. The international questionnaires and evaluation standards mainly used were Life Event Scale (LES), Mini-International Neuropsychiatric Interview (MINI), Childhood Trauma Questionnaire (CTQ) and Patient Health Questionnaire (PHQ-9). PHQ-9 contains 9 common depressive symptoms. Patients were selected and rated based on how they felt. All subjects were informed about the aims and protocols of the data acquisition experiments before EEG recording and all experimental procedures were carried out in accordance with relevant regulations.

### *EEG Recording:*

Most of the current EEG researches use 64 or 128 channels EEG collector. The multi-lead EEG signal acquisition instrument is complicated to operate. Researchers cost half an hour or more time constantly for preparing before signal recorded. Cleaning and applying conductive cream before experiment is a time-consuming job. In the universal environment, it is necessary to reduce the electrode to facilitate the signal acquisition. Given that the prefrontal lobe has a strong correlation with emotional processes and there is no hair interference, a portable EEG acquisition device with only three electrodes (Fp1, Fpz and Fp2) were developed by UAIS from Lanzhou University. The dry electrodes were used to avoid the problem of applying conductive paste. The EEG collection device and acquisition location are shown in Fig. 1.

The experiment began after a minute of relaxation, ninety seconds of eyes-closed and resting-state EEG was recorded. The sampling frequency of the EEG signals was set to 250 Hz with 24 bits A/D convertor precision.

After the completion of all EEG data collection, the data quality was evaluated by the technicians experienced in EEG signal processing. Finally, EEG data of 18 depressed patients and 25 normal subjects were selected, and 40 seconds of each data (10000 points) was taken for calculation in this paper.

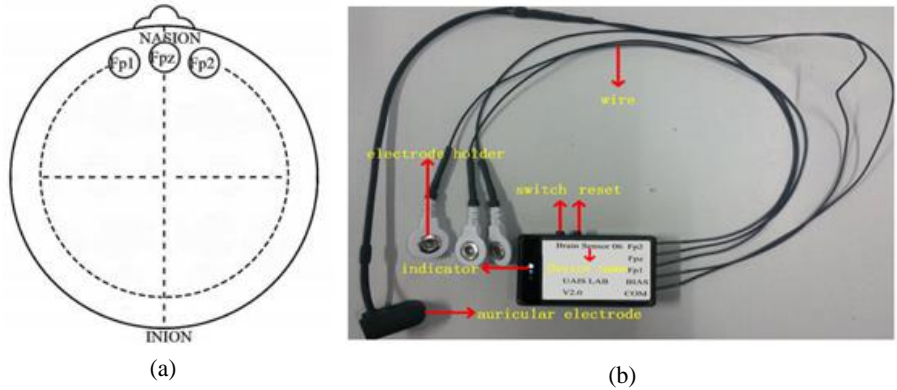

(a) (b)

*Figure 1*. (a) Location of the three electrodes placement. (b) Three-electrode pervasive EEG collection device.

## *Data preprocessing:*

The sampling frequency of the EEG signals was set to 250 Hz. All EEG signals were high-pass filtered with 1 Hz cutoff frequency and low-pass filtered with 40 Hz cutoff frequency. This process not only preserves the signal information to be analyzed, but also avoids the interference of baseline drift and high-frequency electromyography. Then, we used the discrete wavelet transform and Kalman filtering [21] to remove the eye movement.

## *Feature Extraction:*

Based on previous studies, six characteristics of EEG signals were extracted for analysis in this study: three linear features (maximum, mean and center values of the power spectrum) and three non-linear features (correlation dimension, Renyi entropy, and C0 complexity).

(1) Linear features

It has been found that spectral asymmetry is based on the balance between the high and low energy, and the low frequency band of EEG signals seems to be able to represent the EEG of patients with depression [12]. Therefore, the power spectrum features of EEG signals were extracted in order to find the EEG activities spectral asymmetry between depressions and controls.

Through multiple calculations and comparisons, the spectrum features of EEG signals at all signal bands can better represent the depressed EEG than those at delta (1-4 Hz), theta (4-8 Hz), alpha (8-13 Hz) and beta (13-30 Hz) respectively. Hence, for each channel, Welch's method was used to calculate the max, mean and center of the spectrum. Get the linear features matrix with 43 multiple 3 multiple 3.

(2) Nonlinear features

● Correlation Dimension (CD)

CD is a geometric measure of the complexity of the dynamics and can be used to approximate the dimension of a phase space. It is most often computed from the time series diagram constructed from a single data vector. CD reflects the dynamic characteristics of EEG and reflects the correlation degree of EEG sequence itself. The stronger the mental activity is, the higher CD is. When the conditions are fixed, CD is very sensitive to changes in EEG signals and can reflect subtle changes in brain activity.

The basic CD algorithm, the G–P algorithm, was introduced by Grassberger and Procaccia. The delay-time embedding was used to calculate the CD [22]. The value of CD can be computed as follows:

a. EEG data sequence is denoted by x=[x(1),x(2),···,x(N)] with N data points (in this study, N=10000). According to the selection of the time lag τ and embedded dimension m, the phase space vectors X(i) can be reconstructed as:

$$X(i)=\{x(i),x(i+\tau),\cdots,x[i+(m-1)\tau]\}, i=1,2,\cdots N-(m-1)\tau \quad (1)$$

b. Then the correlation integral function C_m (r) can be defined as:

$$C_m(r) = \frac{2}{M(M-1)} \sum_{i \neq j}^{M} \theta[r - |X(i) - X(j)|] \quad (2)$$

Where r is the radial distance around each reference point X(i), M is the number of data points in phase space. |X(i)-X(j)| denotes the Euclid form, and θ(x) is the Heaviside step function which is defined as θ(x)=0 for x<0 and θ(x)=1 for x>0.

c. The CD was defined as:

$$CD = \lim_{r \to 0} \left[ \frac{\ln C(r)}{\ln r} \right] \quad (3)$$

- Renyi Entropy

Kulish et al. [23] begin with the Rényi entropy measure, which does not use sequential properties of the data, but rather is the probability distribution function of the time signal (i.e., the EEG voltages). Renyi entropy is the general form of Shannon entroy, which contains both amplitude and frequency information of the signal. It can be used to analyze the time series of non-stationary process or non-gaussian process. Renyi entropy is an effective time-frequency analysis tool in the study of EEG signals and it can represent the brain' activities complexity [24]. Related research shows that the signal complexity of brain activity in patients with depression is lower significantly than that of controls [25].

The EEG sequence is denoted by x=[x(1),x(2),···,x(N)], 8and x belong to different subintervals according to their amplitude. it was divided into K subintervals. The $p_i$ is corresponding probabilities for each subinterval $\sum_{i=1}^{K} p_i = 1$. Renyi entropy defined as follows [26]:

$$H_\alpha(x) = \frac{1}{1-\alpha} \log \left( \sum_{i=1}^{K} p_i^\alpha \right), \alpha > 0, \alpha \neq 1 \tag{4}$$

- C0 Complexity

Complexity is an indicator of time series. The complexity value indicates the degree of irregularity. C0 complexity is a representation of sequence randomness, which can not only obtain reliable results for short-range data, but also avoid the process of coursing. The main idea of C0 complexity is to decompose the complex event sequence into regular activity and random activity. C0 complexity is defined as the ratio of the area between random activity sequence and the time axis and the area between the whole complex activity sequence and time axis. It is worth mentioning that C0 complexity has the advantage of being less computationally intensive than other quantities which describe complexity. Therefore, we extracted the C0 complexity of EEG signal according to the following process:

Let the time series of EEG signal be x(n)={x(0),x(1),···,x(N-1)},n=0,1,···,N-1, with N sample points. First, the Fast Fourier transform (FFT) [27] of x(n) is calculated:

$$X(k) = \frac{1}{N} \sum_{n=0}^{N-1} x(n) e^{-j(2k\pi n/N)} \tag{5}$$

And, let the mean square value of X(k) be G_N, then:

$$G_N = \frac{1}{N} \sum_{k=0}^{N-1} |X(k)| \tag{6}$$

The X(k) less than or equal to $G_N$ is replaced with zero to get a new spectrum series Y(k):

$$Y(k) = \begin{cases} X(k) & |X(k)|^2 > G_N \\ 0 & |X(k)|^2 \leq G_N \end{cases} \tag{7}$$

Then get y(n) by taking the inverse FFT (IFFT) of Y(k), C0 complexity is defined as:

$$C0 = \frac{\sum_{n=0}^{N-1} |x(n) - y(n)|^2}{\sum_{n=0}^{N-1} |x(n)|^2} \tag{8}$$

Where y(n) is defined as the regular activity part of EEG sequence, and x(n)-y(n) is defined as the random activity part of EEG sequence.

*Category:*

We compared the classification performance of three common classifiers by using the extracted feature set in this study. All classifiers are applied to normalized features. Normalization was performed by subtracting the sample means and dividing by the sample standard deviation, so that the inputs of algorithm have zero means and unit standard deviation.

K-Nearest Neighbor (KNN) was introduced by Dasarathy [28] in 1991, and is a commonly used multivariate classifier. The core idea of KNN is that given a training data set, for the new input, the nearest K (the nearest K neighbors) instances are found in the training data set. The label with the largest number of votes is the label of the new input. This algorithm is widely used in test classification [29], pattern recognition [30], image processing [31] and multi-classification. The advantage of KNN is that the algorithm is simple and easy to implement, which is suitable for sample sets with more overlapping class fields. At the same time, KNN has heavy computation and low prediction accuracy of rare categories when the samples are unbalanced. In this duty, as the parameter of KNN, the K was searched from [1, 2, …, 18].

Logistic regression (LR) [32] is used to deal with the regression problem where the dependent variable is the classification variable. It estimates the class conditional probability using a linear combination of features and logistic regression function. The classification model coefficients are estimated by the LogitBoost algorithm [33] and often used to solve binomial distribution problems.

Support vector machine (SVM) was developed as one of the machine learning methods which are based on statistical theory. The purpose is to find a hyperplane to segment the sample based on the interval maximization, and the problem transformed into a convex quadratic programming [34]. SVM is very efficient in high dimensional space even when the data dimension is larger than the sample size. Using a subset of the training set in decision functions is also memory efficient. It has unique advantages in solving small sample data and nonlinear data. In this study, we use linear kernel functions, soft margin classifier and polynomial optimization algorithm [35] to design SVM model for distinguish the EEG signal depressed or not, and hence, minimize overfitting. Cross-validation and grid method were used to search the optimum values. Due to the small sample size in this study, leave-one-out cross-validation (LOOCV) was used to predict the classification rate, which use only one sample feature set as a test set and the rest as a training set at a time.

The classification experiments were performed in the platform of 3.2 GHz CPU and 8 GB memory and Windows 10 operation system, the SVM was trained using Scikit-learn algorithm package of Python version.

## Results

In this experiment, 18 depressed subjects and 25 normal controls were selected based on the depression scale and data quality assessment. Initial group comparisons between depressed group (age= 31.33±11.30 years) and control group (age=30.28±8.40years), Gender and age were matched between the two groups, and those difference between two groups was not statistically significant ($p<0.05$).

Fig.2 presents the mean and differences between calculated nonlinear features for depressive and control group in each EEG channel (Fp1, Fpz and Fp2). The calculated nonlinear (CD, Renyi entropy and C0 complexity) features are lower in depressive group compared to control group in all channels. However, the reduction of the features with depression is not always statistically significant.

In order to achieve the better model to identify the depressed subjects and normal beings with the three channel EEG, hence, we compared the performance of classification model which were trained based on SVM and KNN. The model training based SVM used all six measures of the three channel (CD, Renyi entropy, C0, maximum, mean and center values of the power spectrum), and the parameters setting in optimal model as follows: the penalty parameter C for the error term is 2, Gaussian kernel was selected and $gamma = 2^{-6}, k = 11.$ However, the optimal model training based KNN used four features (CD, Renyi entropy, C0 and the maximum of the power spectrum) of Fp1, Fpz and Fp2, that is to say k=12. The number of neighbors is 17.

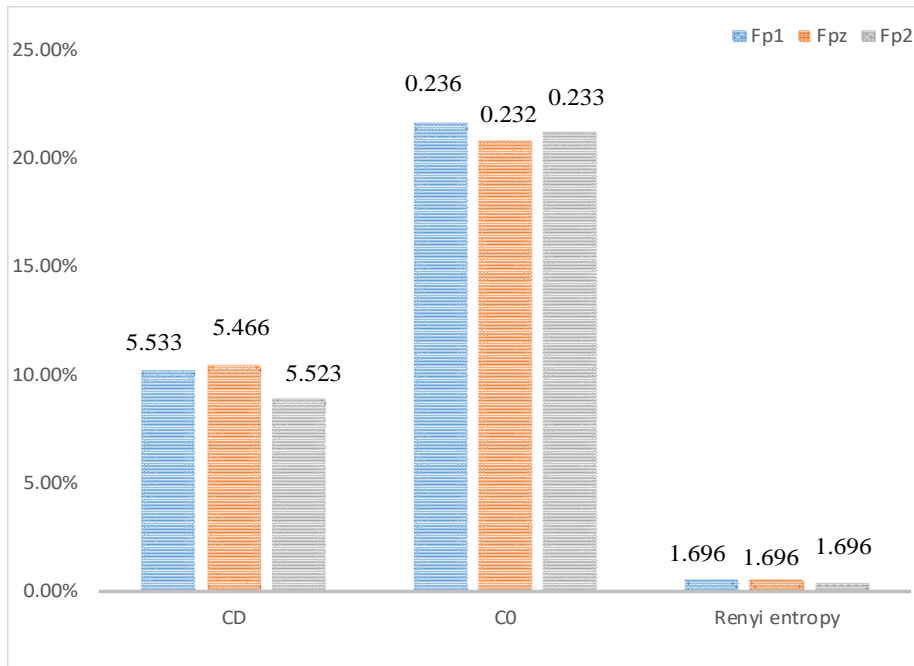

*Figure 2.* Effect of depression on EEG nonlinear (CD, Renyi entropy and C0) measures of three analysed EEG channels as relative difference between values of a measure averaged over all subjects in depressive and control group. The number at the top of the bar shows the mean of measures between all depressions and controls. CD = correlation dimension; C0 = C0 complexity.

Table 1 presents the classification accuracies with different classifiers between depressive subjects and normal controls. The classification accuracy is obtained by the optimal classified model which with the parameters as mentioned above. The results show that the classification accuracy and sensitivity of KNN are 72.25% and 74.06, respectively, which are slightly higher than SVM. However, the specificity of KNN is better than that of KNN.

Table 1. **Classification inputs and results with different classifiers**.

| classifier | features | dimension of features | Acc | Sen | Spe |
|---|---|---|---|---|---|
| KNN | CD, Renyi entropy, C0, Max | 12 | 72.25% | 74.06% | 63.63% |
| SVM | CD, Renyi entropy, C0, Max, Mean, Center | 18 | 71.42% | 72.92% | 67.79% |

The analysis and research in this article is an exploratory attempt. We selected Fp1, Fpz and Fp2 channels which were related to emotion. We have compared six features between 18 depressions and 25 normal controls. The results indicate that the nonlinear features CD, C0 complexity and Renyi entropy with all studied channels in depression were higher than those of controls. Those differences were not statistically significant. This study is established on the basis of the limited samples. According to this method, extending the sample set maybe can obtain the statistically significant differences.

## Discussion

CD is an essential quantitative index in nonlinear EEG analysis. CD is considered to be a reflection of the complexity of the cortical dynamics underlying EEG recordings. Thus a higher CD reflects the increase of theneural activity occurred in brain [36]. C0 complexity is also a measure of the complexity of EEG activity. The more complex and irregular EEG activity is, the higher value of C0 complexity. Renyi entropy reflects the time-frequency information of EEG. A higher Renyi entropy value exhibit a composition complexity of EEG signal. Sunmming up the above, selected nonlinear features mirror the complexity of brain activity. The greater the cognitive load of the brain, the more complex mental activity, and the lager features value. In this study, EEG signal were recorded under the condition with quiet and eyes closed, and the nonlinear features of depressed subjects was higher compared with normal controls. In other words, the brain activities of depressed beings were not as relaxed as that of normal beings although in the quiet closed eyes and relaxed state.

SVM and KNN were compared in our work. SVM have training process and several advantages compared with KNN. The classification result of KNN is slightly higher than SVM in our experiment, it indicates that the feature set and sample size in this study are more suitable for KNN classification.

Although, the Fp1, Fpz and Fp2 channels of EEG signals can be used to make a certain classification of depression and normal beings, there are still some limitations to be mentioned.

First of all, the sample size in this study was too small to make any conclusion for clinical diagnosis. Second, the linear features of EEG signals are rarely discussed in this paper. Effective linear features can help improve the classification accuracy. Finally, the accuracy of classification in this paper are not the best, further research is needed.

In conclusion, our experiment shows the possibility to diagnosis depression using forehead EEG signals. Diagnosis of depression in a pervasive environment, using resting state EEG signal may be more comfortable for depressed subjects.

## Acknowledgments

This work was supported in part by the National Key Research and Development Program of China (Grant No. 2019YFA0706200), in part by the National Natural Science Foundation of China (Grant No.61632014, No.61627808, No.61210010), in part by the National Basic Research Program of China (973 Program, Grant No.2014CB744600), and in part by the Program of Beijing Municipal Science & Technology Commission (Grant No.Z171100000117005).

## References

[1] W. H. Organization, "Depression and other common mental disorders," 2017. [Online]. Available: http://apps.who.int/iris/bitstream/10665/254610/1/WHO-MSD-MER-2017.2-eng.pdf.

[2] W. o. o. M. Health, "Depression: a Global Crisis; World Mental Health Day, October 10, 2012," 2012. [Online]. Available: http://www.who.int/mental_health/management/depression/wfmh_paper_depression_wmhd_2012.pdf.

[3] I. f. H. a. C. Research, "Netherlands Study of Depression and Anxiety (NESDA)," 2018. [Online]. Available: http://www.emgo.nl/research/international-collaborations/longitudinal-cohort-studies/netherlands-study-of-depression-and-anxiety.

[4] Ghorbanian *et al.*, "Exploration of EEG features of Alzheimer's disease using continuous wavelet transform," 2015.

[5] A. F. Leuchter, I. A. Cook, A. M. Hunter, C. Cai, and S. Horvath, "Resting-State Quantitative Electroencephalography Reveals Increased Neurophysiologic Connectivity in Depression," *Plos One,* vol. 7, 2012.

[6] D. Abásolo, R. Hornero, J. Escudero, and P. Espino, "A study on the possible usefulness of detrended fluctuation analysis of the electroencephalogram background activity in Alzheimer's disease," *IEEE Trans. Biomed. Eng.,* vol. 55, no. 9, pp. 2171-2179, 2008.

[7] T. Paus, "Synchronization of neuronal activity in the human primary motor cortex by transcranial magnetic stimulation : an EEG study," *Journal of Neurophysiology,* vol. 4, 2001.

[8] H. Hinrikus *et al.*, "Spectral features of EEG in depression," *Biomedizinische Technik Biomedical Engineering,* vol. 55, no. 3, pp. 155-161, 2010.

[9] A. C. Deslandes *et al.*, "Electroencephalographic frontal asymmetry and depressive symptoms in the elderly," vol. 79, no. 3, pp. 0-322, 2008.

[10] U. R. Acharya, V. K. Sudarshan, H. Adeli, J. Santhosh, J. E. W. Koh, and A. Adeli, "Computer-Aided Diagnosis of Depression Using EEG Signals," *European Neurology,* vol. 73, no. 5-6, pp. 329-336, 2015.

[11] V. A. Grin-Yatsenko, I. Baas, V. A. Ponomarev, and J. D. Kropotov, "Independent component approach to the analysis of EEG recordings at early stages of depressive disorders," vol. 121, no. 3, pp. 0-289, 2010.

[12] V. A. Grin-Yatsenko, I. Baas, V. A. Ponomarev, and J. D. Kropotov, "EEG Power Spectra at Early Stages of Depressive Disorders," *Journal of Clinical Neurophysiology Official Publication of the American Electroencephalographic Society,* vol. 26, no. 6, pp. 401-406, 2009.

[13] V. Knott, C. Mahoney, S. Kennedy, and K. Evans, "EEG power, frequency, asymmetry and coherence in male depression," *Psychiatry Research,* vol. 106, no. 2, pp. 0-140, 2001.

[14] J. J. B. Allen, H. L. Urry, S. K. Hitt, and J. A. Coan, "The stability of resting frontal electroencephalographic asymmetry in depression," *Psychophysiology,* vol. 41, no. 2, pp. 269-280, 2010.

[15] J. Fell, J. R?schke, K. Mann, and C. Sch?ffner, "Discrimination of sleep stages: a comparison between spectral and nonlinear EEG measures," *Electroencephalogr Clin Neurophysiol,* vol. 98, no. 5, pp. 0-410, 1996.

[16] Q. Zhao *et al.*, "An EEG based nonlinearity analysis method for schizophrenia diagnosis," *Biomedical Engineering,* vol. 9, no. 1, p. 136, 2012.

[17] G. M. Bairy, S. Bhat, L. W. J. Eugene, U. C. Niranjan, S. D. Puthankatti, and P. K. Joseph, "Automated Classification of Depression Electroencephalographic Signals Using Discrete Cosine Transform and Nonlinear Dynamics," (in English), *J Med Imag Health In,* vol. 5, no. 3, pp. 635-640, Jun 2015, doi: 10.1166/jmihi.2015.1418.

[18] M. A. Mendez *et al.*, "Complexity analysis of spontaneous brain activity: effects of depression and antidepressant treatment," *Journal of Psychopharmacology,* vol. 26, no. 5, pp. 636-643, 2012.

[19] Y. Li *et al.*, "Abnormal EEG complexity in patients with schizophrenia and depression," *Clinical Neurophysiology Official Journal of the International Federation of Clinical Neurophysiology,* vol. 119, no. 6, pp. 1232-1241, 2008.

[20] M. Bachmann, K. Kalev, A. Suhhova, J. Lass, and H. Hinrikus, *Lempel Ziv Complexity of EEG in Depression*. Springer International Publishing, 2015.

[21] C. Yan, Q. Zhao, B. Hu, J. Li, and P. Hong, "A method of removing Ocular Artifacts from EEG using Discrete Wavelet Transform and Kalman Filtering," in *2016 IEEE International Conference on Bioinformatics and Biomedicine (BIBM)*, 2016.

[22] P. Grassberger and I. Procaccia, "Measuring the strangeness of strange attractors," in *The Theory of Chaotic Attractors*: Springer, 2004, pp. 170-189.

[23] V. Kulish, A. Sourin, and O. Sourina, "Human electroencephalograms seen as fractal time series: Mathematical analysis and visualization," *Computers in Biology & Medicine,* vol. 36, no. 3, pp. 291-302, 2006.

[24] M. A. Colominas, M. E. H. Jomaa, N. Jrad, A. Humeau-Heurtier, and P. Van Bogaert, "Time-Varying Time-Frequency Complexity Measures for Epileptic EEG Data Analysis," (in English), *Ieee T Bio-Med Eng,* vol. 65, no. 8, pp. 1681-1688, Aug 2018, doi: 10.1109/Tbme.2017.2761982.

[25] V. Gaetano, G. R. G., C. Luca, S. E. P., T. C. A, and B. Riccardo, "Nonlinear digital signal processing in mental health: characterization of major depression using instantaneous entropy measures of heartbeat dynamics," *Front Physiol,* vol. 6, 2015.

[26] Michael *et al.*, "Perturbation of differentiated functions during viral infectionin Vivo I. Relationship of lymphocytic choriomeningitis virus and host strains to growth hormone deficiency," 1961.

[27] J. W. Cooley and J. W. Tukey, "An Algorithm for the Machine Calculation of Complex Fourier Series," *Mathematics of Computation,* vol. 19, no. 90, pp. 297-301, 1965.

[28] B. V. Dasarathy, "Nearest neighbor (NN) norms: NN pattern classification techniques," *IEEE Computer Society Tutorial,* 1991.
[29] P. Soucy and G. W. Mineau, "A simple KNN algorithm for text categorization," in *Data Mining, 2001. ICDM 2001, Proceedings IEEE International Conference on*, 2001.
[30] L. Zhang, X. Q. Ding, and X. C. Zhao, "Pattern recognition in near-infrared spectrum by improved knn algorithm," *Modern Electronics Technique,* 2012.
[31] F. Gallegos-Funes, "ABST M-Type K-Nearest Neighbor (ABSTM-KNN) for Image Denoising," *Ieice Transactions on Fundamentals of Electronics Communications & Computer Sciences,* vol. E88-A, no. 3, pp. 798-799, 2005.
[32] D. R. Cox, "The Regression Analysis of Binary Sequences," *Journal of the Royal Statistical Society,* vol. 21, no. 1, pp. 238-238, 1958.
[33] J. Friedman, T. Hastie, and R. Tibshirani, "Additive logistic regression: a statistical view of boosting (with discussion and a rejoinder by the authors)," *The annals of statistics,* vol. 28, no. 2, pp. 337-407, 2000.
[34] C. Cortes and V. N. Vapnik, "Support Vector Networks," *Machine Learning,* vol. 20, no. 3, pp. 273-297, 1995.
[35] J. Piatt, "Fast Training of Support Vector Machines using Sequential Minimal Optimization. Advances in Kernel Methods-Support Vector Learning," ed: MIT Press, Cambridge, Massachusetts, 1998.
[36] Y. J. Lee, Y.-S. Zhu, and Yu-Hong, "The nonlinear dynamical analysis of the EEG in schizophrenia with temporal and spatial embedding dimension," *Journal of Medical Engineering & Technology,* vol. 25, no. 2, pp. 79-83, 2001.